\documentclass{epl}

\title{Dispersion of imbibition fronts} \shorttitle{Imbibition fronts}
\author{Y. M\'elean \inst{1,2} \and D. Broseta \inst{1} \and A. Hasmy \inst{2} \and R. Blossey \inst{3}}
\institute{
  \inst{1} Institut Fran\c{c}ais du P\'etrole, 92852 Rueil-Malmaison
           Cedex, France\\
  \inst{2} Laboratorio de F\'{\i}sica Estad\'{\i}stica de
           Sistemas Desordenados, Centro de F\'{\i}sica, IVIC, Apartado
           21827, Caracas 1020-A, Venezuela\\
  \inst{3} Center for Bioinformatics, Saarland University, P.O. Box 151150, 66041 Saarbr\"ucken, Germany
}


\begin{document}

\maketitle

\begin{abstract}
We have studied the dispersive behaviour of imbibition fronts in a
porous medium by X-ray tomography. Injection velocities were
varied and the porous medium was initially prewetted or not. At
low velocity in the prewetted medium, the imbibition profiles are
found to be distinctly hyperdispersive. The profiles are
anomalously extended when compared to tracer fronts exhibiting
conventional (Gaussian) dispersion. We observe a strong velocity
dependence of the exponent characterizing the divergence of the
dispersion coefficient for low wetting-fluid saturation.
Hyperdispersion is absent at high imbibition velocities or when
the medium is not prewetted.
\end{abstract}


Dispersion is a process by which an initial distribution of fluid
spreads out in a porous medium under the effects of the disordered
nature of the porous structure and the complex flow events
occurring in the pores. This process has practical relevance in
the fields of hydrology (water infiltration, propagation of
pollutants in soils \cite{bear}), petroleum engineering (oil
recovery by water injection \cite{lake}), and biotechnology (gel
and chromatographic analysis \cite{gu}).

A dispersion process can be described on the macroscopic (Darcy)
scale, i.e., for an average fluid content (or saturation) $S$
defined over a ``representative elementary volume" \cite{bear}, by
the equation
\begin{equation} \label{one}
\phi \partial_t S + u \partial_x S = \partial_x (D \partial_x S)
\end{equation}
in which $u$ is the average velocity (or drift), $x$ the position
along the flow direction, and $D$ is the dispersion coefficient.
The simplest dispersion process, referred to as hydrodynamic
dispersion, concerns two fluids of identical nature in which one
contains ``tagged" or tracer molecules. Then, $S$ is the average
tracer concentration, and dispersion arises primarily from the
variation in local fluid velocities (molecular diffusion is
negligible for practical values of $u$). For homogeneous and
non-consolidated porous media, the dispersion coefficient is given
by $ D = u\cdot \ell $, where $ \ell $ is a length of the order of
the typical pore size. Dispersion is Gaussian (i.e., $D$ is {\it
independent} of $S$), and $S$ is a simple function of the reduced
variable $(x-ut/\phi)/\sqrt{t}$ \cite{charlaix}. For many ``real"
porous media, however, one observes a non-Gaussian hydrodynamic
dispersion: the physical mechanisms governing such anomalous
behaviour are being actively investigated \cite{bacri1}.

Here we are concerned with the dispersion behaviour found in
imbibition: a front of wetting fluid is injected through one face
of a horizontally placed porous medium. The medium is initially
filled with a nonwetting fluid. Therefore, a front between wetting
and nonwetting fluid forms which spreads as it advances into the
porous medium, as illustrated in Figure 1.
\begin{figure}[t]
\begin{center}
\includegraphics[scale=0.5]{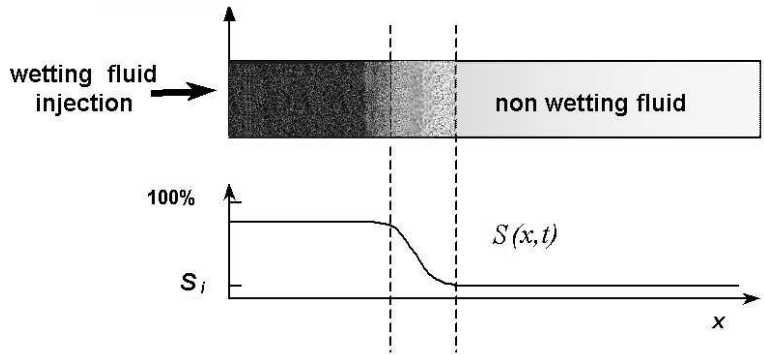}
\end{center}
\caption{Schematic representation of the imbibition experiment.
The wetting phase is injected at a fixed velocity $u$ through the
inlet face of the porous medium, which is closed except at the
outlet. Profiles $S(x,t)$ are obtained by averaging out X-Ray
CT-Scan saturation maps along cross-sections indicated by dashed
lines.
Behind the front, $S(x,t)$ does not
reach 100\% as some nonwetting phase remains trapped.}
\end{figure}
In this case, dispersion is primarily caused by capillary forces
which, depending on pore size and structure, accelerate or slow
down the advance of the wetting fluid: it is usually referred to
as capillary dispersion \cite{davis1,davis2}. Imbibition is
governed by an equation similar to eq.(\ref{one}) in which the
convective term is replaced by $u\partial_xF(S) $, where $F(S)$ is
the (generally nonlinear) fractional flow function from which
dispersion-free solutions are constructed. They are travelling
waves or shocks, depending on the functional form of $F$ and the
initial and boundary conditions \cite{bear,lake}. These solutions
are good approximations of the true solutions when convective
(viscous) forces dominate capillary forces, i.e., for high
capillary numbers.

Anomalous (non-Gaussian) dispersion behaviour has been suggested
theoretically to occur \cite{degennes1,degennes2} and observed
experimentally \cite{bacri2,bacri3} in imbibition at low or
vanishing injection velocity (`passive' imbibition). To
characterize the evolution of the imbibition fronts a power-law
divergence of the dispersion coefficient at low wetting fluid
saturation is used. In the proposed models
\cite{degennes1,degennes2,davis1,davis2} the
(velocity-independent) exponent is related to the pore structure
(through a fractal dimension) and to the range of the
substrate-fluid interactions.

In this letter, we present and discuss a series of imbibition
experiments carried out at different imposed velocities and
initial wetting states of the porous medium. The experiments
consist in monitoring the progression of the front of the wetting
phase (water) displacing a nonwetting phase (oil) by means of an
X-Ray CT-Scanner (CT = computerized tomography). From the measured
water saturation profiles we are able to characterize the
dispersion behaviour in a precise manner, especially at low
injection velocity or capillary number, as we will show. We note
that for our system, we have no a priori knowledge of either the
functional dependence of $F(S)$ nor of the dispersion coefficient
$D$ and its possible dependence on wetting fluid saturation
\cite{remark}.

Our experiments were performed in a porous medium made of
quartzitic grains with radii $\approx 100\, \mu$m, packed in a
cylindrical glass column with a diameter of $3$ cm, and length
$L=40$ cm. The porosity (or void percentage) and absolute
permeability were $\phi \approx 35 \%$, and $k \approx 10\,
\mu$m$^2$ (or Darcy), respectively. Homogeneity was checked from
X-Ray scanner measurements of the dry porous medium. The X-Ray
CT-Scanner device used in this study was a fourth generation
medical model FX/i, General Electric that allow very rapid
measurements: less than 1 minute suffices to measure about 40
images of cross-sections of the porous medium, each cross-section
being placed every cm apart. As the displaced fluid we used
n-decane with 99$\% +$ purity, and the (distilled) water contained
a dopant (30 g/l potassium iodide) to enhance its X-Ray contrast
with the oil. The displacing fluid (water) is slightly more
viscous than the displaced fluid (oil), 1.03 vs. 0.89 $ mPa \cdot
s$, and their interfacial tension at the temperature of the
experiments (23 $^\circ$C) is $\sigma \approx 53$ dyn/cm. The
capillary number, i.e. the ratio of viscous to capillary forces $
Ca = \mu u/ \sigma = \mu Q/\gamma A $ spans more than two orders
of magnitude, from around 10$^{-8}$ to 10$^{-6}$. Here, $Q$ is the
imposed injection rate, and $A$ the cross-sectional area of the
porous medium. The mobility ratio between the two fluid phases is
favorable and one therefore expects the development of a front
which is stable with respect to viscous fingering.

\begin{figure}[t]
\begin{center}
\includegraphics[scale=0.35]{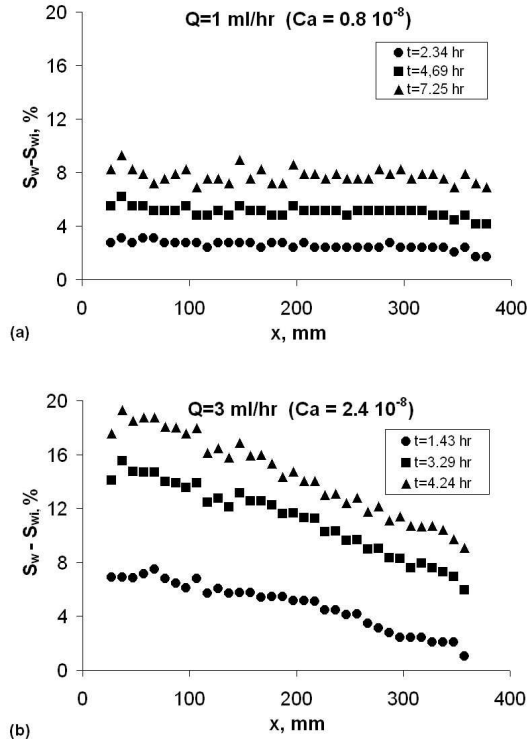}
\end{center}
\caption{Time evolution of wetting saturation profiles from
imbibition experiments in a prewetted porous medium conducted at
values of $Q$ of (a) 1 ml/hr, (b) 3 ml/hr.}
\end{figure}
\begin{figure}[h]
\begin{center}
\includegraphics[scale=0.4]{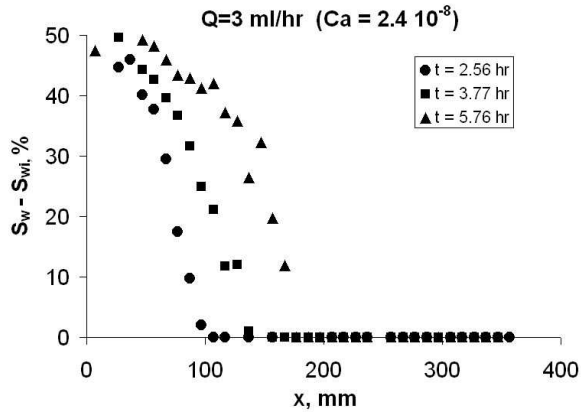}
\end{center}
\caption{Imbibition profiles in a nonprewetted porous medium at
$Q=3$ ml/hr.}
\end{figure}

The procedure for our experiments in a prewetted porous medium is
as follows. First the dry medium is fully saturated with water.
Then, it is flooded at a high rate with an excess of oil, until no
water leaves the medium. Some water cannot be displaced and
remains trapped in the medium. The profile $S_{wi}(x)$ of this
irreducible water measured with the CT-Scanner amounts to 25$\%$
all across the medium, except very near the outlet where it
reaches 45$\%$. Such an accumulation of water is nothing but the
capillary end effect. Then the experiment starts by injecting the
water at an imposed flow rate $Q$. Water saturation profiles
$S_w(x,t)$ are measured at various times $t$ following the start
of the injection. The $S_w(x,t)$ are calculated from the measured
data by linear interpolation with previous CT-Scan data of the
porous medium fully saturated with water or oil.

For an imbibition experiment conducted at a different rate, the
porous medium is again first flooded with oil (as described above)
to reach the state of irreducible water saturation $S_{wi}(x)\sim
25 \%$. Imbibition experiments in a prewetted medium were
conducted at the flow rates $Q=$ 1, 3, 10, 30, 100, and 300 ml/hr,
corresponding to superficial velocities $u/\phi$ (these velocities
would correspond to those of sharp advancing fronts) which equal
$0.14, 0.42, 1.4, 4.2, 14$, and $42$ cm/hr ($Ca$ ranging from
$0.8$ $10^{-8}$ to $2.4$ $10^{-6}$).



We now turn to the description of our findings. The saturation
profiles corresponding to the two lowest velocities in a prewetted
porous medium are shown in Figs. 2. At low injection rates (low
$Ca$) the profiles increase almost uniformly throughout the entire
porous medium. This behaviour qualitatively indicates the presence
of a hyperdispersive imbibition process governed by a dispersion
coefficient which diverges at low saturations as a power law
\cite{degennes1,degennes2,bacri2,bacri3}. This behaviour is
analyzed quantitatively below.

We first assess the role of prewetting on the observed
hyperdispersive behaviour, which we infer from a comparison to a
reference experiment in a non-prewetted porous medium. In this
case, the medium was first fully saturated with oil and then the
water was injected at the imposed velocity. The resulting profiles
are shown in Fig. 3. The saturation front is well-defined and
localized, and the signature of hyperdispersion displayed by the
fronts in Fig. 2 is clearly absent.

By comparison, Figs. 4 a), b) show the saturation profiles in the
medium in the prewetted medium at high velocity. With the
increased velocity (i.e., larger $Ca$) the imbibition fronts
sharpen and become very straight at the two highest rates ($Q=100$
and 300 ml/hr). Such a behaviour might be indicative of a
hypodispersive behaviour, i.e. $D \rightarrow 0 $ for $ S_w
\rightarrow 0 $ \cite{degennes1,degennes2}.

\begin{figure}[t]
\begin{center}
\includegraphics[scale=0.35]{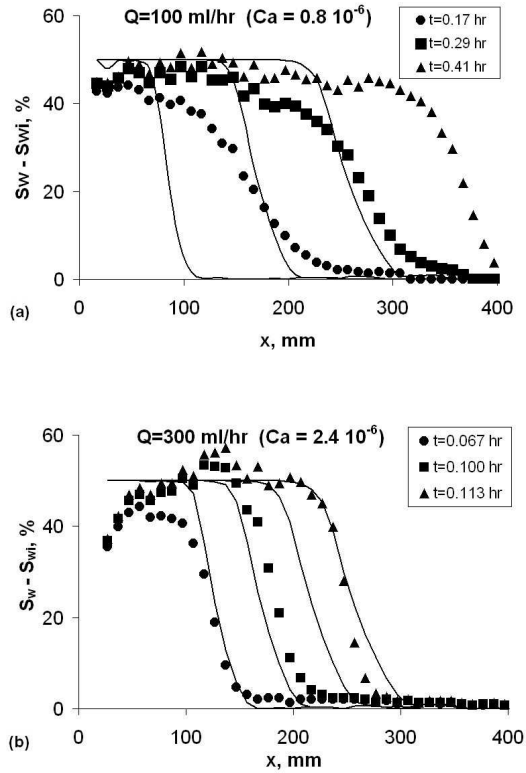}
\end{center}
\caption{Time evolution of wetting saturation profiles from
imbibition experiments conducted at values of $Q$ of (a) 100
ml/hr, (b) 300 ml/hr. Tracer concentration profiles are also
depicted in (a) and (b) by full lines.}
\end{figure}
\begin{figure}[t]
\begin{center}
\includegraphics[scale=0.35]{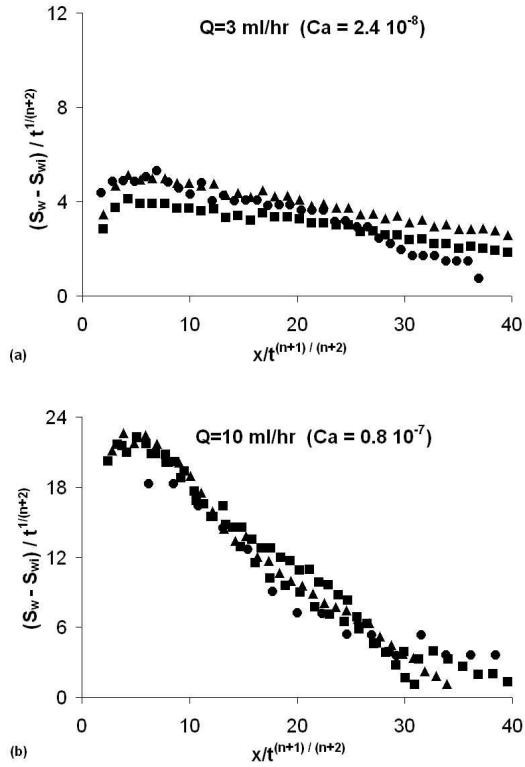}
\end{center}
\caption{Scaled wetting saturation profiles. The corresponding
``best" values of the exponent $n$ are (a) $n=-1.1$, (b) $
n=-0.5$.}
\end{figure}
In order to further characterize the behaviour of these fronts we
have conducted tracer experiments used as a probe for Gaussian
(hydrodynamic) dispersion. In those tracer experiments, the water
doped with the tracer molecules (30 g/l potassium iodide) was
injected at a fixed rate which displaced the pure water initially
fully saturating the porous medium. Then the tracer concentration
profiles were measured at various successive times $t$ with the
X-Ray CT-Scanner. We checked that the tracer profiles could be
superimposed when expressed in terms of the reduced variable
$(x-ut/\phi)/\sqrt{t}$ \cite{bacri1}. In Fig. 4(a) and (b), the
data sets of the tracer fronts are superimposed with the highest
flow rate imbibition fronts. Unfortunately we have too few data to
quantify the capillary dispersion coefficient at low $S_w$ to
unambiguously decide on the presence of hypodispersion in the
fronts. On a longer distance (more than the 40 cm of the porous
medium available to us) we expect the profiles to turn into shocks
\cite{lake}.

We now turn to the quantitative analysis of the saturation
profiles at low flow velocities. In hyperdispersion, the
dispersion coefficient diverges at low wetting phase saturation
according to a power law \cite{degennes1,degennes2}. Early
experiments by Bacri et al. \cite{bacri2,bacri3} had previously
shown evidence for this effect in a prewetted porous medium. In
their case, the data were analyzed assuming $ D \sim S_w^{\, -1}
$, but the acoustic technique used allowed saturation measurements
only for very few positions along the porous medium.

Based on our data, we indeed find that the saturation profiles can
be interpreted in terms of a power law divergence of the
dispersion coefficient, however of the form
\begin{equation}
D(S_w) \sim (S_w-S_{wi})^n\, , n < 0
\end{equation}
in which case eq.(1) has a self-similar scaling solution for the
saturation profile for small $u$ \cite{bacri3,remark,mayer}.
Consequently, the saturation profiles superimpose when scaled by
$t^{1/(n+2)}$ if positions are rescaled by $t^{(n+1)/(n+2)}$. By
using this scaling solution, we can collapse our low flow rate
saturation data onto a single master curve (Fig. 5).

However, the exponent $n$ allowing the best collapse exhibits a
strong dependence with the imposed flow rate: $n\approx -1.1$ for
$Q=3$ ml/hr, and $n \approx -0.5$ for $Q= 10$ ml/hr. For the
lowest rate $Q=1$ ml/hr, where a ``giant hyperdispersivity" is
found characterized by an almost homogeneous profile spanning the
whole system (see Fig. 2 a), we could not determine a scaling
exponent.

These results have not been anticipated theoretically. In Refs.
\cite{degennes1,degennes2} it was argued that a divergence of $D$
should be related in a simple scaling fashion to the
fluid-substrate molecular interactions that determine the
thickness of wetting films, or to surface geometry. Our porous
medium does not possess a complex (fractal) pore geometry
\cite{davis1,davis2}; in addition, since the wetting fluid
occupies about 25$\%$ of the pore space, the thickness of the
wetting layers is much larger than the range (several tens of
nanometers) of the fluid-substrate interactions. In our
experiments it is rather capillarity that controls the
distribution of the fluid in the pores space. Most of the wetting
phase is in the ``bulk" form of menisci and ``pendular"
structures, sometimes referred to as layers \cite{blunt2},
occupying corners and substrate concavities. These wetting layers
provide a continuous and conductive pathway across the porous
medium, and presumably swell rather uniformly across the porous
medium at low injection velocities.

A challenge for future studies is to understand how the flow
events on the pore scale give rise to the observed hyperdispersive
behaviour on the macroscopic (Darcy) scale. These events are
complex, as they include cooperative pore invasion, flow in layers
along the substrate micro-roughness, and collapse between wetting
layers. A very useful tool for modeling these events is provided
by pore-network models that have shed some light on various
patterns and macroscopic flow parameters observed in imbibition
under variation of capillary number, contact angle and wetting
phase saturation \cite{blunt1,blunt2}. However, so far these
models do not account for a variation in the conductance of the
wetting layers (i.e., their swelling) which we believe is needed
to reproduce the strong hyperdispersive effect we observe. Some
progress in pore-network modeling is to be expected from the
incorporation of proper swelling rules, e.g. for layers in corners
\cite{weislogel}.

To conclude, we have demonstrated experimentally, for the first
time, that the presence of wetting layers is needed for
hyperdispersion to occur in imbibition in a porous medium. For low
wetting-phase saturation the dispersion coefficient behaves like a
power-law with a velocity-dependent effective exponent. New
theoretical advances are needed to account for these results which
we expect can serve as a benchmark for future pore-network
modeling studies.

\acknowledgments We are grateful to C. Fichen and C. Schlitter for
their assistance in the CT-Scan measurements, and to O. Vizika and
B. Noetinger for discussions. Y.M. was supported by the
Post-Graduate Cooperation Program (PCP) France-Venezuela. R.B. was
supported by DFG under Schwerpunktprogramm ``Wetting and Structure
Formation at Surfaces" (Bl-256) and the Flemish Government under
Grant (VIS/97/01). We further acknowledge support under TOURNESOL
grant T2000.013.


\begin{thebibliography}{0}
\bibitem{remark} Within the standard Darcy formalism for two-phase
flow in porous media, $ D $ and $F$ can be expressed in terms of
the capillary pressure and relative permeabilities, which are
phenomenological functions of $S$ (see refs.
\cite{bear,lake,degennes1,degennes2,bacri2,bacri3}).
\end{thebibliography}
\end{document}